\documentclass[sigconf]{acmart}
\AtBeginDocument{%
  \providecommand\BibTeX{{%
    \normalfont B\kern-0.5em{\scshape i\kern-0.25em b}\kern-0.8em\TeX}}}

\setcopyright{acmlicensed}
\copyrightyear{2024}
\acmYear{2024}
\acmDOI{10.1145/3649921.3650011}

\copyrightyear{2024}
\acmYear{2024}

\acmConference[FDG 2024]{Proceedings of the 19th International Conference on the Foundations of Digital Games}{May 21--24, 2024}{Worcester, MA, USA}

\acmBooktitle{Proceedings of the 19th International Conference on the Foundations of Digital Games (FDG 2024), May 21--24, 2024, Worcester, MA, USA}
\acmISBN{979-8-4007-0955-5/24/05}

\begin{document}

\title{The NES Video-Music Database: A Dataset of Symbolic Video Game Music Paired with Gameplay Videos}

\author{Igor Cardoso}
\affiliation{%
  \department{Departamento de Informática}
  \institution{Universidade Federal de Viçosa}
  \streetaddress{Av. P.H. Rolfs}
  \city{Viçosa}
  \state{Minas Gerais}
  \country{Brazil}}

\author{Rubens O. Moraes}
\affiliation{%
  \department{Departamento de Informática}
  \institution{Universidade Federal de Viçosa}
  \streetaddress{Av. P.H. Rolfs}
  \city{Viçosa}
  \state{Minas Gerais}
  \country{Brazil}}

\author{Lucas N. Ferreira}
\affiliation{%
  \department{Departamento de Informática}
  \institution{Universidade Federal de Viçosa}
  \streetaddress{Av. P.H. Rolfs}
  \city{Viçosa}
  \state{Minas Gerais}
  \country{Brazil}}


\begin{abstract}
Neural models are one of the most popular approaches for music generation, yet there aren't standard large datasets tailored for learning music directly from game data. To address this research gap, we introduce a novel dataset named NES-VMDB, containing 98,940 gameplay videos from 389 NES games, each paired with its original soundtrack in symbolic format (MIDI). NES-VMDB is built upon the Nintendo Entertainment System Music Database (NES-MDB), encompassing 5,278 music pieces from 397 NES games. Our approach involves collecting long-play videos for 389 games of the original dataset, slicing them into 15-second-long clips, and extracting the audio from each clip. Subsequently, we apply an audio fingerprinting algorithm (similar to Shazam) to automatically identify the corresponding piece in the NES-MDB dataset. Additionally, we introduce a baseline method based on the Controllable Music Transformer to generate NES music conditioned on gameplay clips. We evaluated this approach with objective metrics, and the results showed that the conditional CMT improves musical structural quality when compared to its unconditional counterpart. Moreover, we used a neural classifier to predict the game genre of the generated pieces. Results showed that the CMT generator can learn correlations between gameplay videos and game genres, but further research has to be conducted to achieve human-level performance.
\end{abstract}





\maketitle

\section{Introduction}



Indie game developers typically handle every aspect of their game themselves, from coding to visual assets. However, music and sound effects are usually outsourced to third-party producers or sourced from online resources (paid or open). While outsourcing audio production might be a viable option for established indie developers, it's less accessible for those with hard budget constraints, particularly in developing countries. To make music production for games more democratic, one could train generative models capable of composing music based on game data. These models could help in many project phases, from quickly including background music in early prototypes to guiding the composition of the final soundtrack.

Given that neural generative models are one of the most popular methods in music generation \cite{wang2020learning}, one could train one of these models (e.g., an auto-regressive language model) to compose music from game data. However, to the best of our knowledge, large datasets of video game music paired with corresponding game data are currently unavailable. In this paper, we present the NES Video-Music Database (NES-VMDB)\footnote{https://github.com/rubensolv/NES-VMDB}, comprising 98,940 gameplay videos from 389 NES (Nintendo Entertainment System) games, each video paired with its respective background music in symbolic format.

The NES-VMDB is an extension of the Nintendo Entertainment System Music Database (NES-MDB), designed for constructing automatic music composition systems for the NES audio synthesizer \cite{donahue2018nes}. The NES-MDB dataset comprises 5,278 music pieces from 397 NES games. The NES-VMDB associates 4,070 of these pieces with short gameplay clips from the game scenes where these musical compositions are played. For example, the Super Mario Bros. World 1-1 music theme is paired with multiple clips of a player navigating World 1-1. 

We use gameplay videos, instead of other game data (e.g., tilemaps), because videos are game-independent data formats that capture sufficient information to support music composition appropriately. In other words, gameplay videos do not rely on the internal data representations of the game. Our objective is to enable music generation models that allow users to input a brief gameplay clip of their own game and receive background music that complements the scene pictured in the clip. We envision these generators being utilized by indie game developers to generate music in different stages of their projects, from musical sketches at early stages to final soundtracks at later versions of the game.

To pair the NES-MDB MIDI pieces with gameplay clips, we initially obtained long-play videos from YouTube for 389 NES-MDB games. A long-play video is a play-through from beginning to end with no audio comments. Subsequently, we divided each video into 15-second-long clips and isolated their audio. Additionally, we synthesized all MIDI files from the NES-MDB using the NES synth provided alongside the dataset\footnote{\url{https://github.com/chrisdonahue/nesmdb}}. Finally, we employed a fingerprinting algorithm to retrieve, for each 15-second audio clip, the most similar piece among the synthesized ones. The MIDI file associated with the retrieved piece was then matched with its corresponding video query.

In addition to the dataset, we established a baseline generator based on the Controllable Music Transformer (CMT) \cite{di2021video}, a state-of-the-art model for general music composition conditioned on video inputs. CMT is conditioned during inference with rhythmic features extracted from an input video. We trained it with the NES-VMDB MIDI pieces and then generated new music by conditioning it with rhythmic features extracted from gameplay clips. We compared the outputs of this baseline against an unconditional CMT. 

We used objective music structure metrics to compare the quality of the music produced by these methods. Results showed that the conditional pieces have a structure more similar to human-composed pieces than the unconditional ones. We also trained a classifier based on the CMT architecture to predict the game genre of the generated pieces. Results showed that the conditional CMT was able to learn correlations between gameplay videos and game genre, but qualitatively its pieces are still far from human-composed ones. Thus, there are many research opportunities to improve upon this baseline.

\section{Related Work}


This work is primarily related to the NES-MDB dataset \cite{donahue2018nes}, symbolic music generation from videos, and music generation for video games. In this section, we review the most relevant methods from each of these areas.

\subsection{The NES-MDB dataset}

\begin{figure}[!t]
    \centering
    \includegraphics[width=8.5cm]{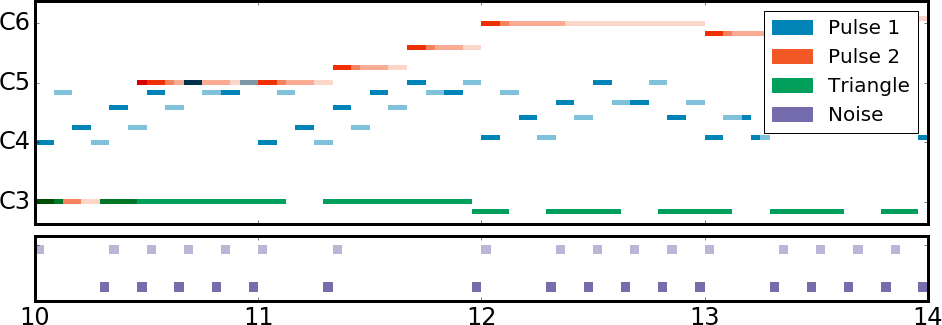}
    \caption{The piano roll representation of a snippet of a piece from the NES-MDB dataset. The horizontal axis represents time and the vertical axis represents pitch. The color of the notes represents their channel \cite{donahue2018nes}. }
    \label{fig:nesmdb}
\end{figure}

The NES-MDB dataset compiles 5,278 music pieces from the soundtracks of 397 NES games. Each piece features four out of the five instrument voices of the NES synthesizer: two pulse-wave generators (P1, P2), a triangle-wave generator (TR), and a percussive noise generator (NO). The fifth voice, an audio sample playback channel, was excluded for simplicity. Figure \ref{fig:nesmdb} illustrates the piano roll representation of a piece snippet from the NES-MDB dataset.

All pieces were extracted from the assembly code of NES games, ensuring precise timings and parameter values necessary for an accurate reproduction of the music as it sounds in the NES system. The dataset provides the extracted pieces in various representations, including MIDI, music score, and VGM. It encompasses compositions from 296 unique composers, featuring over two million notes in total.

\subsection{Symbolic Music Generation from Videos}

Neural models for generating symbolic music conditioned on videos have garnered increasing interest in recent years. One common challenge involves generating a monophonic composition for a given instrument based on the movements of a musician playing that instrument. For instance, \citet{su2020audeo} employed a convolutional neural network to encode a video of piano players and a GAN to generate a piano piece in piano roll format. \citet{gan2020foley} adopted a Transformer architecture with an encoder that processes a video of a musician playing an instrument (e.g., piano, bassoon, cello, etc.) and a decoder that generates a symbolic monophonic composition matching the movements in the performance. \citet{su2020multi} investigated a similar problem but with a VQ-VAE model that generates compositions directly as an audio signal instead of a symbolic sequence.

Another related problem is generating polyphonic music for a general video. For instance, \citet{di2021video} introduced a transformer-based approach called Controllable Music Transformer (CMT). CMT extracts music features from MIDI files to train a music language model. During inference,  rhythmic features are replaced with those from a given video to enable controllable generation. An interesting aspect of this work is that it does not require a dataset of videos paired with music. \citet{zhuo2023video} proposed an alternative approach called V-MusProd, which learns mappings from video to music using a dataset of video-music pairs. As part of this work, they introduced a dataset called SymMV.

\subsection{Symbolic Music Generation for Video Games}

Various approaches have been proposed for generating music in the context of video games. For instance, \citet{williams2015dynamic} employed a rule-based system to generate soundtracks by transforming pre-generated melodies, aligning them with the emotional context of annotated game scenes. \citet{scirea2017affective} used evolutionary algorithms in MetaCompose, a framework designed to generate real-time background music for games using an evolutionary algorithm. \citet{cardinale2023harmonymapper} designed a system called HarmonyMapper combining the MAP Elites algorithm with Neo-Riemannian music theory to generate diverse chord sequences in terms of the emotions they aim to evoke. \citet{ferreira_ismir_2019} introduced a dataset of piano arrangements for video game soundtracks and a learning method to compose game music with a specified sentiment. \citet{ferreira2020computer, ferreira2022controlling} extended this work by incorporating search-based decoding methods to enhance the quality and emotional content of the generated pieces.

\section{The NES-VMDB Database}

Our goal with the NES-VMDB database is to enable the composition of background music for games from gameplay clips. The initial step in creating our dataset was to exclude any MIDI file from the NES-MDB dataset that doesn't represent a music file, such as sound effects. To accomplish this, we defined a complete piece as any MIDI file with a duration greater than 8 seconds. We selected 8 seconds because it corresponds to the length of one phrase, comprising 4 bars at 120 bpm. In essence, we considered a file a piece if it contained at least 4 bars of music at 120 bpm. This cleaning step eliminated 1,208 MIDI files, leaving a total of 4,070 pieces.

After completing the cleaning step, we searched on YouTube for long-play videos corresponding to each of the 397 games in the NES-MDB dataset. For each game, our search query was constructed as follows: ``\{GAME\_NAME\} NES World of Longplay''. World of Longplay\footnote{\url{https://www.youtube.com/channel/UCVi6ofFy7QyJJrZ9l0-fwbQ}} is a YouTube channel that hosts long-play videos for various gaming platforms, including consoles, arcades, PC, etc. Currently, they feature 1,083 videos of NES games, encompassing licensed and unlicensed titles released in Japan, Europe, and the USA. Our search yielded a video with 360p resolution for 389 games, with long-play videos unavailable for only 8 games. Since our query explicitly specified ``World of Longplay'', most videos were retrieved from this channel. For games where YouTube didn't find a video in this channel, but in others, we used these extra resources to compile as many videos as possible. All clips sum up to a total of 474 hours of video. Table \ref{tab:yt_channels} lists the top 5 channels by the number of retrieved videos.

\begin{table}[!h]
    \centering
    \caption{Number of videos retrieved by YouTube channel. The Other category includes 54 channels, 3 with 3 games, 6 with 2 games, and 45 with 1 games.}
    \label{tab:yt_channels}
    \begin{tabular}{lc}
        \toprule
       \bf{YouTube Channel}  & \bf{Number of Videos} \\ \midrule
        World of Longplay & 303 \\ 
        NintendoComplete & 7 \\ 
        Nenriki Gaming Channel & 6 \\ 
        30-30 Club & 4 \\ 
        Ice Jacket & 3 \\ 
        Other &  66\\ \bottomrule
    \end{tabular}
\end{table}

\begin{figure}[!t]
    \centering
    \includegraphics[width=8.5cm]{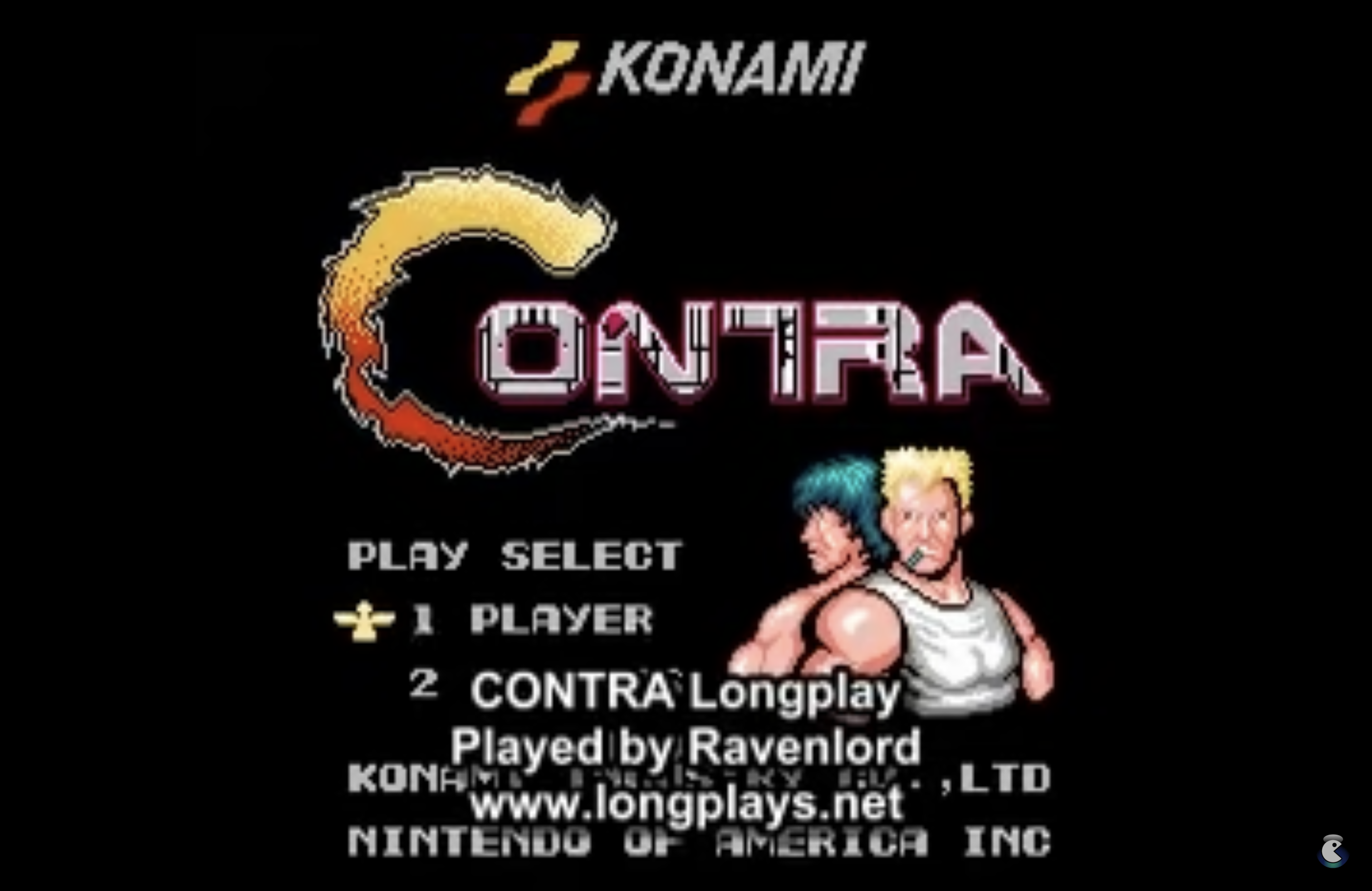}
    \caption{One of the first frames of the Contra long-play from the World of Longplay channel.}
    \label{fig:wol}
\end{figure}

In all channels, the videos consist of direct screen captures from an NES emulator. They have no voice-overs or edits, except for the World of Longplay videos, which have text labels with metadata and the channel icon on top of the first few frames. Figure \ref{fig:wol} illustrates an initial frame from the Contra long-play on the World of Longplay channel.

After downloading all long-play videos, we divided each one into non-overlapping clips of $15$ seconds. This segmentation yielded 98,940 short clips, with an average of $225.32$ clips per game (standard deviation of $501.39$). We then separated the audio tracks from each clip. While these clips exclusively contain game audio, they frequently mix background music and sound effects. Moreover, some clips may capture moments of scene transitions, potentially including the end of the piece from the first scene and the beginning of the piece from the second scene. After the segmentation, we synthesized each of the 4,070 music pieces from the NES-MDB using the NES synthesizer provided with the dataset. These synthesized music pieces are crucial to mapping the clips to NES-MDB MIDI files.

We employed an audio fingerprinting algorithm to automatically associate the gameplay clips with their corresponding MIDI files. It is worth highlighting that manually pairing the clips with the MIDI files would be prohibitively time-consuming, given the large volume of game clips and music pieces we have in our dataset (474 hours of video and 4,070 music pieces). We used Dejavu\footnote{\url{https://github.com/worldveil/dejavu}}, an open-source implementation of the Shazam algorithm \cite{wang2003industrial} as our fingerprinting algorithm. Dejavu transforms an audio clip into a distinctive acoustic fingerprint, comprising multiple hash values that capture its unique characteristics. To generate a fingerprint, it initially segments the audio signal into short, overlapping time frames and computes the corresponding spectrogram. It then identifies peaks in the spectrogram, representing distinctive frequency-time pairs, and hashes this information into a condensed code, forming the unique fingerprint. These fingerprints, consisting of multiple hash values, are subsequently stored in a database. To retrieve an entry from the database, Dejavu repeats this process for an audio query and matches its fingerprint against the stored ones.

To facilitate retrieval, we create a separate fingerprint database for each game. Consequently, when provided with a 15-second audio clip as a query, Dejavu only compared it against the fingerprints of pieces from the game to which the query belongs. To create the NES-VMDB dataset, we generated a query for each of the 98,940 audio clips and used Dejavu to retrieve the synthesized music piece with the highest number of fingerprint matches. 

Subsequently, we paired the MIDI file used for synthesizing the retrieved piece with the video clip associated with that query. It is noteworthy that Dejavu, like Shazam, exhibits robustness against noise \cite{wang2003industrial}. Therefore, even if our queries contain sound effects, they can accurately retrieve the correct music piece. To quantify Dejavu's performance, we sampled 30 query results at random and manually evaluated them. Dejavu retrieved 25 results correctly and 1 incorrectly. The other 4 results were audio clips containing only sound effects that Dejavu tried to fit with the best match from their respective game. 



\section{Experiments}

\begin{table*}[t]
    \centering
        \caption{Objective comparison between CMT Conditioned, CMT Unconditioned, and Human-composed pieces.}
    \label{tab:objetive}
    \begin{tabular}{lccc}
         \toprule
         \bf{Metric}  & \bf{CMT Conditioned} & \bf{CMT Unconditioned} & \bf{Human} \\ \midrule
        \bf{Grooving Pattern Similarity}   & 0.821       & 0.694           & 0.999    \\
        \bf{Number of Unique Pitch Classes}      & 9.840       & 9.043           & 10.755   \\
        \bf{Pitch Class Histogram Entropy} & 2.703       & 2.590           & 2.970    \\
        \bf{Pitch Range}          & 41.160      & 37.457          & 50.085   \\
        \bf{Number of Notes Played Concurrently }             & 1.311       & 1.214           & 2.055 \\ \bottomrule
    \end{tabular}

\end{table*}

Our primary goal with the NES-VMDB dataset is to support generative models that learn a mapping from videos to music. To facilitate future research, we introduce a baseline method based on the Controllable Music Transformer (CMT) \citet{di2021video}. We trained our CMT model with the MIDI files of the NES-VMDB as a music language model (i.e., predicting the next token given a symbolic music context). We first augmented the dataset following the methodology of \citet{oore2017learning}. We transposed each piece to every key, increased and decreased the tempo by 10\%, and adjusted the velocity of all notes by 10\%. After augmentation, we encoded all these pieces with the CMT encoding scheme. CMT quantizes a MIDI file into sixteenth-note timesteps and, for each timestep, generates a vector with 7 music features: token type (rhythmic or melodic), timestep type (beat or bar), density, strength, instrument, pitch, and duration.

We defined our CMT model with 8 transformer blocks each containing 8 attention heads. The model's hidden and feed-forward inner layer sizes are 256 and 2,048, respectively. The dropout rate in each layer is set to 10\%. The input sequence length is padded to 10,000 with the \emph{<EOS>} (End of Song) token. We trained this model with all augmented pieces for 25 epochs using the Adam optimizer and a learning rate of 1e-4, which took approximately 3 days on a single NVIDIA GeForce RTX 4070 Ti with 12GB of memory.

\begin{figure}[t]
    \centering
    \includegraphics[width=8.5cm]{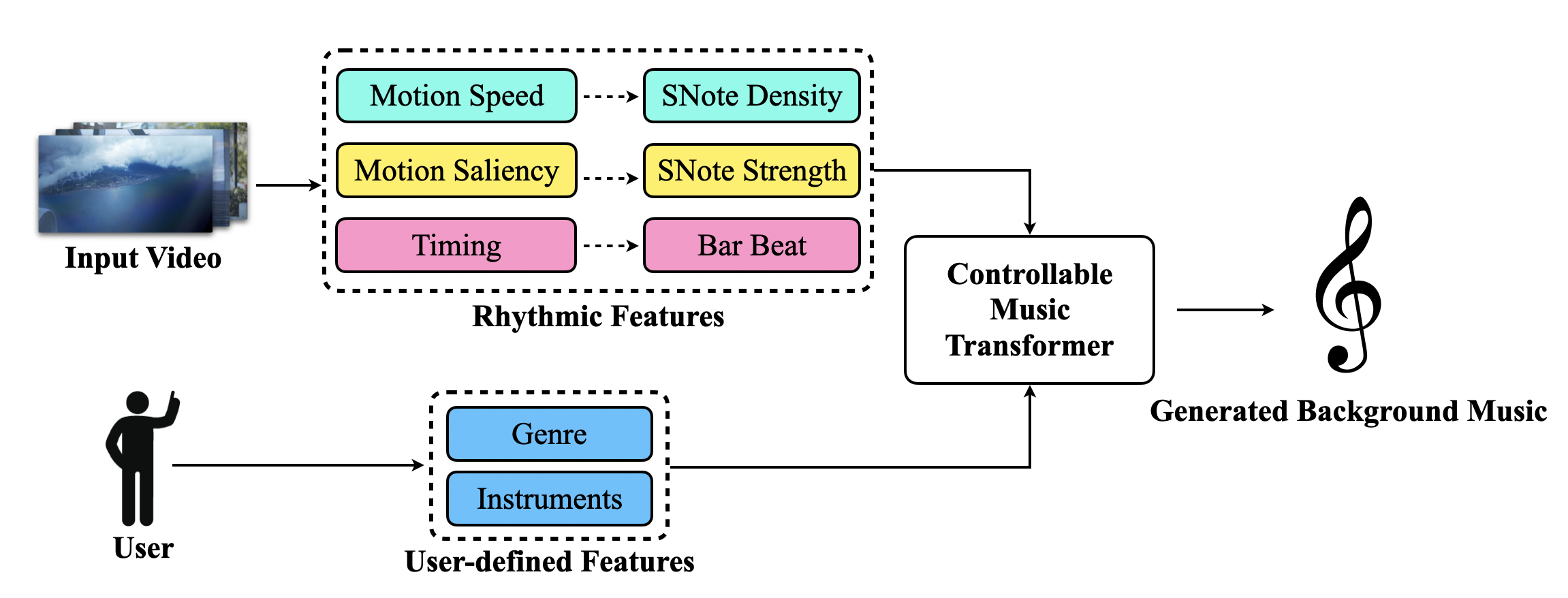}
    \caption{CMT approach to condition music generation from input videos \cite{di2021video}.}
    \label{fig:cmt}
\end{figure}

After training our CMT model, we randomly selected 28 games from the NES-VMDB dataset, and for each game, we selected 5 clips at random. We then generated a piece for each clip, yielding 140 conditioned pieces. To condition the generation, CMT replaces the strength and density attributes of the generated tokens during inference with strength and density values extracted from the input video, as shown in Figure \ref{fig:cmt}. It is important to highlight that while CMT allows users to control genre and instruments, we didn't use these attributes.

We evaluate this conditioning approach against an unconditional one and human-composed pieces. Thus, we used CMT to generate 140 unconditioned pieces to represent the unconditioned method. Moreover, we selected 140 ground truth MIDI pieces (as given by Dejavu) to represent the human method. These pieces came from the same gameplay clips we used to generate the conditioned pieces.

\subsection{Music Structure Metrics}
\label{sec:structure}

We compared these pieces using five objective metrics calculated with the MusPy \cite{dong2020muspy} toolkit: Pitch Class Histogram Entropy, Grooving Pattern Similarity, Pitch Range, Number of Unique Pitch Classes, and Number of Notes Being Played Concurrently. All metrics represent features that can be extracted from symbolic music and are commonly used to compare generated music to human compositions \cite{dong2018musegan}. Specifically, Grooving Pattern Similarity and Pitch Class Histogram Entropy were employed to evaluate the CMT model in its original paper \cite{di2021video}. The former is the mean Hamming distance of neighboring measures and helps in measuring the music’s rhythmicity. The latter is the Shannon entropy of the normalized note pitch class histogram and helps assess the music’s quality in tonality. The Number of Unique Pitch Classes and Pitch Range are extra metrics to evaluate tonality, and the Number of Notes Played Concurrently helps evaluate harmony quality.

These metrics are useful for measuring the distance between generated pieces and human-composed ones. Thus, the closer the values are to those of the human pieces, the better. Table \ref{tab:objetive} reports the average results for these metrics. Overall, the conditioned CMT outperformed the unconditional one in all metrics, indicating that the approach proposed by \citet{di2021video} generates musical structures closer to human-composed pieces than an unconditional method. The Grooving Pattern Similarity is the metric in which the conditioned CMT model had the lowest distance, indicating that its generated pieces are more closely related to the human ones in rhythm than harmony or melody. 

The low distance in the Number of Notes Being Played Concurrently suggests that harmonically the conditioned pieces are not too far from human pieces as well. In terms of melody, the low distances in Pitch Class Histogram Entropy and Number of Unique Pitch Classes but high in the Number of Notes Played Concurrently suggest that while the conditioned pieces have a similar relative variation (entropy) and usage of pitch classes, they do not explore different registers of a given pitch as much as human pieces.

\subsection{Game Genre Classification}

To evaluate whether the generated pieces align with the game genres of the clips they were conditioned on, we trained a neural classifier to predict the game genre from a symbolic music piece. For the classifier training, we categorized all 389 NES-VMDB games by genre. We initially extracted the genre from the right panel of each game's Wikipedia article. If games featured multiple genres, we kept only the first. This process yielded 40 specific genres, such as ``Block breaker'', ``Vehicular combat'', and ``Carnival''. This space of classes is considerably large for the amount of data we have. Thus, we reduced this initial list to a more manageable one.

To compile a shorter list of broader genres we searched for the most common genres across the XBOX, PlayStation, Nintendo Switch, and Steam online game stores. We found the following 11 genres as a result of this process: Shooters, Sports, Platformers, RPG, Puzzle, Action, Fighting, Strategy, Simulation, Adventure, and Racing. After compiling this broader list, we employed ChatGPT (version 3.5) to map the 40 specific Wikipedia genres to the 11 general ones, using the prompt ``Map each specific game genre in List A to a more general genre in List B''. Table \ref{tab:wiki_genre_mapping} presents the results of the mapping generated by ChatGPT.

\begin{table}[t]
    \centering
    \caption{Mapping performed by ChatGPT from genres retrieved from Wikipedia to genres listed in online games stores.}
    \label{tab:wiki_genre_mapping}
    \begin{tabular}{ll}
    \toprule
    \textbf{Wikipedia Genres} & \textbf{Stores Genres} \\
    \midrule
    Scrolling shooter & Shooters \\
    Rail shooter & Shooters \\
    2D action platformer & Platform \\
    Run and gun & Shooters \\
    Block breaker & Puzzle \\
    Puzzle-platform & Puzzle \\
    Beat 'em up & Fighting \\
    Multi-directional shooter & Shooters \\
    Turn-based strategy & Strategy \\
    Run-and-gun & Shooters \\
    Maze & Puzzle \\
    Casino & Simulation \\
    Action-adventure & Adventure \\
    Platform-adventure & Adventure \\
    Science fiction & Adventure \\
    Side-scrolling action & Action \\
    Shoot 'em up & Shooters \\
    Light gun shooter & Shooters \\
    Fixed shooter & Shooters \\
    Action RPG & RPG \\
    First-person rail shooter & Shooters \\
    Vehicular combat & Action \\
    Graphic adventure & Adventure \\
    Hack and slash & Action \\
    Action adventure & Adventure \\
    Tactical role-playing & RPG \\
    Pinball & Simulation \\
    Baseball & Sports \\
    Arcade style racing & Racing \\
    Side-scrolling & Action \\
    Rail shooter & Shooters \\
    Carnival & Simulation \\
    Modern first-person adventure & Adventure \\
    Educational & Simulation \\
    Tile-matching & Puzzle \\
    Action platformer & Platform \\
    Children's book & Adventure \\
    Shooting gallery & Shooters \\
    Multidirectional shooter & Shooters \\
    Business simulation & Simulation \\
    \bottomrule
    \end{tabular}
\end{table}

\begin{figure}[b]
    \centering
    \includegraphics[width=8.5cm]{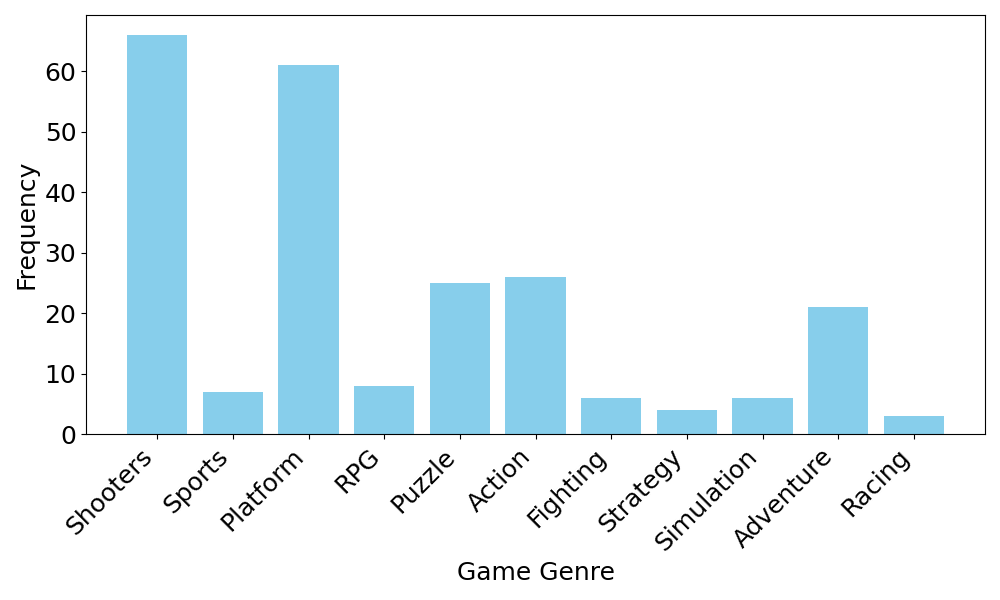}
    \caption{Distribution of genres in the NES-VMDB database.}
    \label{fig:genres}
\end{figure}

We labeled the 389 games in our dataset with this mapping given by ChatGPT. Figure \ref{fig:genres} shows the distribution of examples per genre. Shooter and Platform are the most prominent genres with approximately 60 games each. Puzzle, Action, and Adventure are the second most prominent, with approximately 20 games each. All the other genres have less than 10 examples each.

Our genre classifier has a similar architecture to the CMT model we used to generate conditional and unconditional music. The main difference is that the genre classifier has 4 transformer blocks instead of 8. All the other hyperparameters of the model were set the same. The classifier was trained with the Adam optimizer for 10 epochs with a learning rate of 1e-5.

We adhered to a data split similar to the original NES-MDB dataset: 80\% for training and 20\% for testing, ensuring no composer appeared in two subsets. Originally, the NES-MDB dataset separates 10\% of the data for validation, but given that our game music genre dataset is relatively small for its number of classes, we use the validation data for testing (we didn't perform a validation step). The classifier achieved a test accuracy of $29\%$, which is approximately $3.2$ times better than a random guess ($\approx 9\%$). This result shows that predicting the game genre of an NES music piece is a difficult task. In other words, when listening to different NES pieces, it is hard to identify the game genre only by the game music. This might be because many different genres used the same style of music during the NES generation.

We used our classifier to predict the genre of the 140 conditioned and 140 human pieces generated to evaluate music structural quality (see Section \ref{sec:structure}). It correctly predicted 22\% and 34\% of the genres for the conditioned and human pieces, respectively. The genre accuracy of the conditioned CMT pieces is 12\% lower than the human-composed ones. This result suggests that the conditioned CMT generator can learn correlations between gameplay videos and game music genres. However, both the music generator and the genre classifier can be considerably improved to produce results similar to human performance.

\section{Conclusions and Future Work}

This paper presented the NES-VMDB dataset, a collection of 98,940 NES gameplay videos paired with their background music in symbolic format (MIDI). Our goal with this dataset is to support new generative models that map gameplay videos to game music in symbolic format. We focus on symbolic music, as opposed to audio, because we envision these future learning algorithms to be employed as part of game development pipelines, especially within the indie community. Thus, symbolic music (e.g., MIDI) has the advantage of being easily editable, while audio files are harder to manipulate. In other words, the generated melodies, harmonies, and rhythms won't necessarily be realized acoustically with the NES synth. Developers can use the generator to prototype soundtracks with different instrumentation that are inspired by NES music but not direct variations of it.

The videos in our dataset were retrieved from YouTube and the music was from a previous dataset called NES-MDB. We paired the videos with the pieces automatically using an audio fingerprinting algorithm similar to Shazam. We extracted the audio signal from the video and used it as a query to retrieve the most similar piece from a database of fingerprints constructed for each game in NES-MDB.

Additionally to the dataset, we trained a Controllable Music Transformer \cite{di2021video} as a first baseline for generating NES music conditioned on gameplay videos. We evaluated this approach against its unconditional version by generating a set of 140 pieces and computing objective music structure metrics. Results showed that the conditional model can generate music structurally more similar to human-composed pieces than pieces generated unconditionally. Moreover, we labeled all games in our dataset according to their genre and trained a CMT classifier to predict the genre of a given music piece. We used this classifier as a proxy for human evaluators, labeling the genre of a set of generated pieces. Results showed that the CMT generator can learn correlations between gameplay videos and game genres.

While we achieved positive results with conditional CMT, the problem of generating game music from videos isn't solved. First, the accuracy of the genre classifier can be improved to evaluate the generated pieces better. Moreover, when listening to many generated pieces, especially long ones, one can identify structural issues such as excessive repetition or lack of musical form that could be addressed by a new model. Thus, as future work, we plan to build an end-to-end model to generate NES music from gameplay clips. We also plan to improve the genre classifier by fine-tuning a pre-trained music model \cite{zeng2021musicbert}.

\begin{acks}
We would like to express our gratitude to all the YouTube creators for their dedication and effort in producing high-quality long-play videos of the NES games. Special thanks go to those behind the World of Longplay, who produced most of the data we used.
\end{acks}



\bibliographystyle{ACM-Reference-Format}
\bibliography{main}










\end{document}